\documentclass[letterpaper,titlepage,11pt]{article}
\usepackage{hyperref}

\usepackage{amssymb,amsmath,amsfonts}
\usepackage{epsfig}

\def\clock{{\count0=\time
           \divide\count0 60
           \ifnum\count0<10 0\fi\the\count0
           \multiply\count0 -60 \advance\count0 \time
           :\ifnum\count0<10 0\fi \the\count0
         }}
\newcommand{\timestamp}{{\small\vbox{\hbox{\tt\jobname.tex}
\hbox{\the\day/\the\month/\the\year, \clock}}}}

\newcommand{\be}{\begin{equation}} \newcommand{\ee}{\end{equation}}
\newcommand{\bea}{\begin{eqnarray}} \newcommand{\eea}{\end{eqnarray}}

\newcommand{\id}{\hbox{1\kern-.27em l}}
\newcommand{\sid}{\hbox{\scriptsize1\kern-.27em l}}

\newcommand{\we}{\kern-.1em\wedge\kern-.1em}
\newcommand{\scal}{\kern-.13em\cdot\kern-.13em}

\newcommand{\II}{I\kern-.09em I}

\newcommand{\Z}{\mathbb{Z}} 
\newcommand{\R}{\mathbb{R}}

\newcommand{\nn}{\nonumber}

\newcommand{\spa}{\ , \ \ }

\numberwithin{equation}{section}

\begin{document}

\begin{titlepage}

\rightline{\vbox{\small\hbox{\tt hep-th/0309116} }}
\vskip 3cm

\centerline{\Large \bf New Phase Diagram for}
\vskip 0.1cm
\centerline{\Large \bf Black Holes and Strings on Cylinders}

\vskip 1.6cm
\centerline{\bf Troels Harmark$^{\rm a,b}$
and Niels A. Obers$^{\rm b}$
}
\vskip 0.5cm
\centerline{{}$^{\rm a}$\sl Jefferson Physical Laboratory}
\centerline{\sl Harvard University}
\centerline{\sl Cambridge, MA 02138, USA}

\vskip 0.7cm

\centerline{{}$^{\rm b}$\sl The Niels Bohr Institute}
\centerline{\sl Blegdamsvej 17, 2100 Copenhagen \O, Denmark}

\vskip 0.5cm

\centerline{\small\tt harmark@nbi.dk, obers@nbi.dk}

\vskip 1.6cm

\centerline{\bf Abstract}
\vskip 0.2cm
\noindent
We introduce a novel type of phase diagram for black holes and
black strings on cylinders. The phase diagram involves a new
asymptotic quantity called the relative binding energy. We plot
the uniform string and the non-uniform string solutions in this new
phase diagram using data of Wiseman. Intersection rules for branches of
solutions in the phase diagram are  deduced from a new Smarr formula that
we derive.


\end{titlepage}


\pagestyle{plain}
\setcounter{page}{1}

\section{Introduction}
Neutral and static
black holes and black strings on cylinders $\R^{d-1} \times S^1$
have proven to have a
very rich phase structure, a structure which is still largely unknown.
Gregory and Laflamme \cite{Gregory:1993vy,Gregory:1994bj} discovered
that uniform black strings on cylinders,
i.e. strings that are symmetric around the cylinder, are unstable
to linear perturbation when the mass of the string
is below a certain critical mass.
This was interpreted to mean that a light uniform string decays
to a black hole on a cylinder since that has higher entropy.

However, Horowitz and Maeda \cite{Horowitz:2001cz} argued
that such a transition has an intermediate step: The light uniform
string decays to a non-uniform string, which then eventually decays
to a black hole.
This prompted a search for this missing link, and a branch of
non-uniform string solutions was found in
\cite{Gubser:2001ac,Wiseman:2002zc}.%
\footnote{This branch of non-uniform string solutions was
in fact already discovered by Gregory and Laflamme
in \cite{Gregory:1988nb}.}
Furthermore, in \cite{Choptuik:2003qd} the classical decay
of the Gregory-Laflamme instability was studied explicitly.

A related question is what happens to a black hole on a cylinder
when one increases the mass. Obviously the black hole must have a phase
transition into a black string when the black hole horizon
becomes so big that it cannot fit on the cylinder.
Several suggestions for the phase structure of
the Gregory-Laflamme instability, the black hole/string transitions
and the uniform/non-uniform string transitions have been put forward
\cite{Horowitz:2001cz,Gubser:2001ac,Harmark:2002tr,Horowitz:2002dc,Kol:2002xz,Wiseman:2002ti,Harmark:2003fz,Kol:2003ja}
but it is unclear which of these, if any, are correct.%
\footnote{Other recent and related work includes
\cite{Casadio:2000py,Casadio:2001dc,Horowitz:2002ym,DeSmet:2002fv,Kol:2002dr,Sorkin:2002nu,Frolov:2003kd,Emparan:2003sy}.}

In this paper we give new tools to understand the phase
structure of neutral and static black holes and strings on cylinders.
As part of this, we introduce a new type of phase
diagram for black holes and strings on cylinders.
The phase diagram
has two parameters, the mass $M$ and a new asymptotic quantity
called the relative binding energy $n$.
We explain why this phase diagram is natural to use from several
points of view, and how it is useful for
study of thermodynamics.
We plot moreover the uniform string branch and
the non-uniform string branch of Wiseman \cite{Wiseman:2002zc}
in the phase diagram.

In Section \ref{secdefMn} we consider the linearized Einstein equations
for a localized distribution of static and neutral matter on a cylinder
$\R^{d-1} \times S^1$. In particular we introduce binding energy along
the periodic direction.
From this we define the total mass $M$
and the so-called relative binding energy $n$. We furthermore define the
Fourier modes of the mass-distribution.
We explain that these quantities completely characterize the energy
momentum tensor.
Finally, we show how to measure these quantities from the metric
independent of the gauge.
In particular,
this enables us to find $M$ and $n$ for any black hole and string
solution on the cylinder from the asymptotics of the metric.

In Section \ref{secvar} we introduce a new phase diagram for
static and neutral black holes/strings with the mass $M$ and
relative binding energy $n$ as the variables. We plot the three
known branches of solutions in this phase diagram. This includes
the black hole branch and the uniform string branch, which we
explain has $n=1/(d-2)$. We furthermore plot the non-uniform
string branch obtained numerically by Wiseman
\cite{Wiseman:2002zc} for $d=5$ in the $(M,n)$ phase diagram.

 A new Smarr formula
for static and neutral black holes/strings on cylinders
\begin{equation}
\label{smarrform0} TS = \frac{d-2-n}{d-1} M \ ,
\end{equation}
is derived in Section \ref{secbranch}. We furthermore discuss how
this relation can be used to study the thermodynamics of branches
of solutions in the $(M,n)$ phase diagram. In particular, one of
the consequences of the Smarr formula is an intersection rule for
branches of solutions, stating which branch has the highest
entropy.

In Section \ref{secvirt} we expand on the virtues of the $(M,n)$
phase diagram. We also compare to another type of phase diagram,
the $(\lambda,M)$ diagram, that has been used in the literature so
far. In particular, some important differences are pointed out.

We have the discussion and conclusions in Section \ref{secconcl}.
Among other things,
we discuss the uniqueness of black hole/string solutions on a
cylinder.

A number of appendices is included providing some more details and
derivations.
In Appendix \ref{appgreen} we derive the Green function
for the Laplacian on a cylinder $\R^{d-1} \times S^1$.
In Appendix \ref{appMGL} we list for completeness the
Gregory-Laflamme masses for uniform black strings in $5 \leq D
\leq 10$ space-time dimensions. Appendix \ref{appwis} provides
various details of the numerically obtained non-uniform solution
of Wiseman \cite{Wiseman:2002zc}, which we used to plot this
branch in our new phase diagram.

\subsubsection*{Note added}

After this paper originally was submitted the paper \cite{Kol:2003if} 
appeared which overlaps with some of the considerations
of this paper.

\section{Definition of mass $M$ and relative binding energy $n$
on cylinders}
\label{secdefMn}

It is well known that for static and neutral mass distributions in
flat space $\R^d$ the leading correction to the metric
at infinity is given by the mass.
In this section we show that on a cylinder $\R^{d-1} \times S^1$
we instead need {\sl two} independent asymptotic quantities
to characterize the leading correction to the metric at infinity.
In the following we give two such quantities, one of which is the
mass $M$, and show how they can be measured asymptotically on the cylinder.
We also determine the other independent asymptotic quantities that
one can define.

We note that similar asymptotic quantities as the ones given
here have previously been considered 
in \cite{Traschen:2001pb,Townsend:2001rg}.

We consider a static and neutral distribution of matter
which is localized on a cylinder $\R^{d-1} \times S^1$.
We denote here the time-coordinate as $t = x^0$, the compact direction
as $z = x^d$ and the directions of $\R^{d-1}$ as $x^1,...,x^{d-1}$.
The direction $z$ is periodic with period $2\pi R_T$.
We also define the radial coordinate $r$ by
$r^2 = (x^1)^2 + \cdots + (x^{d-1})^2$.
With this, we assume the energy momentum tensor
\begin{equation}
T_{00} = \varrho
\spa
T_{zz} = - b \ ,
\end{equation}
with all other components being zero.
Here $\varrho$ is the mass density and $b$ is the binding energy
or negative pressure along the $z$ direction.
Note that the conservation of the energy momentum tensor
then imposes that $\partial_z b = 0$.
We assume in the following that $\varrho$ and $b$ are functions
of $r$ and $z$ only, i.e. that the matter is spherically symmetric
on $\R^{d-1}$.

We define then the mass $M$ and the {\sl relative binding energy} $n$ as
\begin{equation}
M = \int d^{d} x \, \rho(x)
\spa
n = \frac{1}{M} \int d^{d} x \, b(x) \ .
\end{equation}
$n$ is called the relative binding energy since it is the
binding energy per unit mass.
We furthermore define Newton's gravitational potential as
\begin{equation}
\nabla^2 \Phi = 8\pi G_{\rm N} \frac{d-2}{d-1} \varrho \spa
\nabla^2 = \frac{\partial^2}{\partial z^2}
+ \frac{\partial^2}{\partial r^2} + \frac{d-2}{r} \frac{\partial}{\partial r}
\ .
\end{equation}
Away from the localized distribution of mass
we can then write%
\footnote{Notice that we assume $\rho(r,-z)=\rho(r,z)$. This we restrict
ourselves to since it applies to all cases considered in Section
\ref{secvar}.}
\begin{equation}
\label{thephi}
\Phi (r,z) = - \frac{d-2}{(d-1)(d-3)} \frac{8\pi G_{\rm N}}{2\pi R_T}
\sum_{k=0}^\infty \frac{1}{r^{d-3}} \, h \! \left( \frac{kr}{R_T} \right)
\cos \! \left( \frac{kz}{R_T} \right) \varrho_k \ ,
\end{equation}
with the Fourier modes
\begin{equation}
\varrho_k = \left\{
\begin{array}{ll}
\int_0^{2\pi R_T} dz \int_0^\infty dr \, r^{d-2} \rho (r,z) &
\ , \ k=0 \\
2 \int_0^{2\pi R_T} dz \int_0^\infty dr \, r^{d-2} \,
g \! \left(\frac{kr}{R_T}\right) \cos \! 
\left( \frac{kz}{R_T} \right) \rho (r,z) &
\ , \ k \geq 1 \end{array} \right.
\end{equation}
where
\begin{equation}
g(x) = \Gamma\left( \frac{d-1}{2} \right) 2^{\frac{d-3}{2}}
x^{-\frac{d-3}{2}} I_{\frac{d-3}{2}}(x)
\spa
h(x) = \frac{2^{-\frac{d-5}{2}}}{\Gamma\left(\frac{d-3}{2}\right)}
x^{\frac{d-3}{2}} K_{\frac{d-3}{2}}(x) \ .
\end{equation}
Here $I_s(x)$ and $K_s(x)$ are
the modified Bessel functions of the second kind.
Note that $g(0)=h(0)=1$. See Appendix \ref{appgreen} for
the Green function of the Laplacian on a cylinder $\R^{d-1} \times S^1$ from which
the above can be derived.

We see that the Fourier modes
$\varrho_k$ characterize the distribution of mass,
and that in particular $\varrho_0 = M/\Omega_{d-2}$.
Since $\partial_z b = 0$ we see that $b(r,z)$ is completely characterized
by $M$ and $n$.
Therefore, if we know $M$, $n$ and $\varrho_k$, $k \geq 1$, we know
the energy momentum tensor completely
(when we are away from the mass distribution).
In accordance with this we show in the following
that $M$, $n$ and $\varrho_k$ precisely
correspond to the gauge invariant information for a given metric.

The Einstein equations in a $(d+1)$-dimensional space-time are
\begin{equation}
\label{einst}
R_{\mu \nu} - \frac{1}{2} g_{\mu \nu} R = 8 \pi G_{\rm N} T_{\mu \nu}
\spa \mu,\nu = 0,1,...,d \ .
\end{equation}
We consider now these equations for a weak gravitational field
\begin{equation}
g_{\mu \nu} = \eta_{\mu \nu} + h_{\mu \nu} \spa
|h_{\mu \nu} | \ll 1 \ ,
\end{equation}
where $\eta_{\mu \nu} = \mbox{diag}(-1,1,...,1)$ is the Minkowski metric
and $h_{\mu \nu}$ is the correction to the Minkowski metric.%
\footnote{We define $h^{\mu \nu} = \eta^{\mu \rho} \eta^{\nu \sigma}
h_{\rho \sigma}$ so that $g^{\mu \nu} = \eta^{\mu \nu} - h^{\mu\nu}$.}
One can consider $h_{\mu \nu}$ as the first order correction
to the metric when making an expansion of the metric
in powers of Newton's constant $G_{\rm N}$.
For use in the following,
note that our assumption that the mass distribution is static means that
$\partial_t g_{\mu \nu} = 0$ and $g_{0i} = 0$, $i =1,...,d$
and hence  $R_{0i} = 0$.

We now impose the {\sl harmonic gauge} for $h_{\mu \nu}$
\begin{equation}
\partial^\nu h_{\mu \nu} = \frac{1}{2} \partial_\mu ( \eta^{\rho \sigma}
h_{\rho \sigma} ) \ .
\end{equation}
With this condition, the Ricci tensor is to first order in $G_{\rm N}$ given by
\begin{equation}
R_{\mu \nu} = - \frac{1}{2} \nabla^2 h_{\mu \nu}\ .
\end{equation}
The Einstein equations \eqref{einst} thus becomes
\begin{equation}
\label{heqs}
\nabla^2 h_{00} = - \frac{16\pi G_{\rm N}}{d-1}
\left( (d-2) \varrho - b \right)
\spa
\nabla^2 h_{zz} = - \frac{16\pi G_{\rm N} }{d-1}
\left( \varrho - (d-2) b \right) \ ,
\end{equation}
\begin{equation}
\nn
\nabla^2 h_{ij} = - \delta_{ij} \frac{16\pi G_{\rm N} }{d-1}
\left( \varrho+b \right) \ .
\end{equation}
Define now the potential $B$ for the binding energy by
\begin{equation}
\nabla^2 B = -\frac{8\pi G_{\rm N}}{d-1} b \ .
\end{equation}
Away from the localized distribution of mass (where $b(x)$ is also zero)
we have then
\begin{equation}
B (r) =  \frac{1}{(d-1)(d-3)}
\frac{8\pi G_{\rm N}}{\Omega_{d-2} 2\pi R_T} \frac{nM}{r^{d-3}} \ .
\end{equation}
The unique solution of the Einstein equations \eqref{heqs}
is therefore
\begin{equation}
h_{00} = - 2 \Phi - 2 B
\spa
h_{zz} = - \frac{2}{d-2} \Phi - 2(d-2) B
\spa
h_{ij} = \delta_{ij} \left( - \frac{2}{d-2} \Phi + 2 B\right) \ .
\end{equation}
We can thus conclude that $h_{\mu \nu}$ is completely determined in
the harmonic gauge by $M$, $n$ and $\varrho_k$, $k \geq 1$.
Obviously this means then that there are
no additional gauge invariant parameters than $M$, $n$ and $\varrho_k$.

We now consider how to measure $M$, $n$ and $\varrho_k$ when
we do not assume $h_{\mu \nu}$ to be in the harmonic gauge.
To analyze this one can try to make coordinate transformations
that take us into other gauges.
However, we first need to define what coordinate
transformation we allow, i.e. which general boundary conditions we
impose.

Consider now the $(r,z)$ coordinate system along with the
first order correction $h_{\mu \nu}$ to the metric without imposing
a specific gauge.
We first impose the condition
that when $G_{\rm N} \rightarrow 0$ the zeroth order
metric should be $-dt^2 + dz^2 + dr^2 + r^2 d\Omega_{d-2}^2$.
Moreover, when $r\rightarrow \infty$ the metric should also
reduce to $-dt^2 + dz^2 + dr^2 + r^2 d\Omega_{d-2}^2$.
We impose furthermore that $z$ is a periodic coordinate with
period $2\pi R_T$.
Finally, we impose that the leading part of $h_{\mu \nu}$
for $r \rightarrow \infty$ is independent of $z$. This is
physically reasonable to impose since very far away from the
mass distribution the true physical dependence on the periodic direction $z$
should diminish just like it does
in \eqref{thephi} where the leading term is proportional to
$1/r^{d-3}$.
Clearly this condition is satisfied in the harmonic gauge.

Thus, if we consider an allowed coordinate transformation
from $(r,z)$ to $(\tilde{r},\tilde{z})$, we see that
it has to be of order $G_{\rm N}$ (or higher),
that $\tilde{r}/r \rightarrow 1$ and $\tilde{z}/z \rightarrow 1$
for $r \rightarrow \infty$, and that $\tilde{z}$ should
be periodic with period $2\pi R_T$. Finally, the leading part
of the transformed
$h_{\mu \nu}$ for $\tilde{r} \rightarrow \infty$
should be independent of $\tilde{z}$.

That any allowed coordinate transformation should be of order $G_N$,
immediately has the consequence that $h_{00}$ is invariant
under coordinate transformations, i.e. gauge invariant.

We have in addition one more gauge invariant piece in the metric.
Since $h_{\mu \nu}$ is independent of $z$ we can easily compute
that $R_{zz} = - \frac{1}{2} \nabla^2 h_{zz}$.
Therefore, in any choice of gauge we have that the leading contribution
to $h_{zz}$ for large $r$ is of the form
$h_{zz} = c_z / r^{d-3} + \cdots$.
If one wishes to change the constant $c_z$ by a coordinate transformation,
one needs to define a new coordinate as
$\tilde{z} = z ( 1 + u/r^{d-3} ) + \cdots $, which gives
$h_{\tilde{z}\tilde{z}} = (c_z - 2u)/r^{d-3} + \cdots$.
However, clearly the coordinate $\tilde{z}$ would not be a periodic
coordinate with constant period $2\pi R_T$ anymore so this is
not an allowed coordinate transformation.
Therefore, $c_z$ is invariant
under the allowed coordinate transformations.

In conclusion, we have shown that
\begin{equation}
g_{00} = -  1 - 2 \left( \Phi + B \right) \ ,
\end{equation}
to first order in $G_{\rm N}$, and that the leading correction
to $g_{zz}$ for $r\rightarrow \infty$ is
\begin{equation}
g_{zz} = 1 + \frac{1}{(d-1)(d-3)}
\frac{16\pi G_{\rm N}}{\Omega_{d-2} 2\pi R_T} \left[ 1 - (d-2) n \right]
 \frac{M}{r^{d-3}} \ ,
\end{equation}
independently of the choice of gauge. Using this we can measure
$M$, $n$ and $\varrho_k$, $k\geq 1$, from a given metric.

In particular, if one wishes to measure the mass $M$ and
the relative binding energy $n$ from a given
metric which has the leading behavior
\begin{equation}
\label{ctcz}
g_{00} = - 1 + \frac{c_t}{r^{d-3}} \spa
g_{zz} = 1 + \frac{c_z}{r^{d-3}} \spa
\end{equation}
for $r\rightarrow \infty$, then $M$ and $n$ are given by
\begin{equation}
\label{findMn}
M = \frac{\Omega_{d-2} 2\pi R_T}{16 \pi G_N} \left[ (d-2) c_t - c_z \right]
\spa
n = \frac{c_t - (d-2) c_z}{(d-2) c_t - c_z} \ .
\end{equation}
These formulas enables us to read off $M$ and $n$ for any
static and neutral solution.

Note that from $M$ and $n$
we can define the {\sl tension} in the circle direction
as
\begin{equation}
\mu = \frac{nM}{L} \ ,
\end{equation}
where $L = 2\pi R_T$ is the circumference of the circle.
The tension $\mu$ has previously been considered 
in \cite{Traschen:2001pb,Townsend:2001rg,Traschen:2003jm,Shiromizu:2003gc}.

\subsubsection*{Bounds on the relative binding energy}

We have the following general bound on the relative binding energy
\begin{equation}
0 \leq n \leq d-2 \ ,
\end{equation}
which holds for any static and neutral solution.

The bound $n \geq 0$ was found in \cite{Traschen:2003jm,Shiromizu:2003gc}. 
The proof is similar
to the simple proof of the Positive Energy Theorem 
of \cite{Witten:1981mf} which was generalized to include horizons
in \cite{Gibbons:1983jg}.
One of the basic assumptions that goes into the proof is
the Dominant Energy Condition.

The bound $n \leq d-2$ follows instead from the Strong Energy Condition.
The Strong Energy Condition gives that 
$T_{00} + \frac{1}{d-1} T^\mu_{\ \mu} \geq 0$ 
which in the above analysis
of Newtonian matter gives $c_t \geq 0$ from which $n \leq d-2$ follows.
Basically the Strong Energy Condition imposes in this case
that gravity is not repulsive asymptotically.

\section{$M$ and $n$ as variables for phase
diagram}
\label{secvar}

In this section we consider static and neutral
black holes and strings on a cylinder.
We have shown above that for
any neutral and static object on a cylinder
$\R^{d-1} \times S^1$ we can measure the mass $M$ and the relative
binding energy $n$.
It is therefore natural to use these two independent quantities as
the variables in a two-dimensional phase diagram for static and
neutral black holes and strings
on a cylinder. As we review below
we need a two-dimensional phase diagram to illustrate the known solutions.

\begin{figure}[ht]
\centerline{\epsfig{file=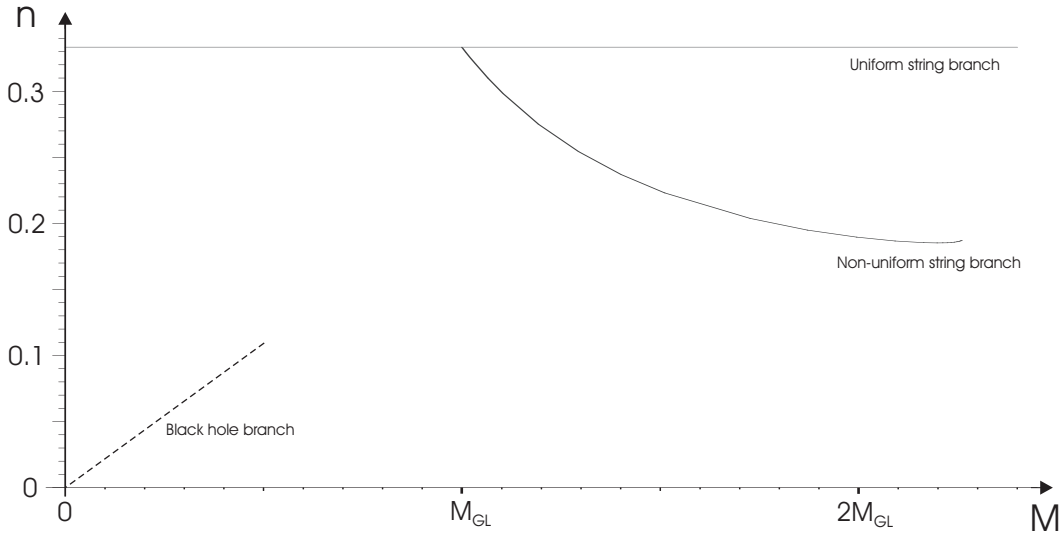,width=14cm,height=7cm}}
\caption{$(M,n)$ phase diagram for $d=5$ containing
the uniform string branch and the non-uniform string branch
of Wiseman. The black hole branch is sketched.}
\label{phase1}
\end{figure}

We have three known branches of solutions corresponding to
neutral and static black holes and strings on a cylinder.
First we have the black hole branch. This branch starts at $(M,n)=(0,0)$,
since it is physically clear that a
small mass black hole will have negligible relative binding energy
(see \cite{Harmark:2002tr} for quantitative discussion of this).
The black hole branch should terminate at a certain critical mass
since the horizon radius at some point becomes too large to fit on the
cylinder.
This branch is the subject of \cite{Harmark:2002tr},
where an ansatz for solutions in this branch was given.
We have
made a qualitative illustration of this branch in the phase diagram
in Figure \ref{phase1}.

The second branch is given by the uniform neutral black string solutions,
which exist for any mass $M$. These solutions are obtained as
a direct product of a Schwarzschild black hole solution with a circle,
i.e. the metric is
\begin{equation}
\label{metunstr}
ds^2 = -f dt^2 +  f^{-1} d r^2  +  dz^2 +
 r^2 d \Omega_{d-2}^2  \spa f = 1 - \left( \frac{r_0}{r} \right)^{d-3} \ .
\end{equation}
From the definition \eqref{ctcz} we infer that $c_z = 0$ for any $M$,
so that it follows from \eqref{findMn}  that $n = 1/(d-2)$
for all $M$. We also illustrated this in the phase diagram in
Figure \ref{phase1}
(we are considering the special case $d=5$ in the diagram).
As was discovered by Gregory and Laflamme in \cite{Gregory:1993vy},
the solutions in this branch are
classically stable when $M > M_{GL}$ and classically
unstable when $M < M_{GL}$, where $M_{GL}$ is found numerically
in \cite{Gregory:1993vy} (in Appendix \ref{appMGL} we have listed
the Gregory-Laflamme masses for $d=4,..,9$).

The third branch is the one discovered by
\cite{Gregory:1988nb,Gubser:2001ac}
and further explored
in \cite{Wiseman:2002zc,Wiseman:2002ti}.
This branch of solutions begins with the
uniform black string solution in $(M,n)=(M_{GL},1/(d-2))$
and continues with increasing $M$ and $n$ (at least for the known cases
$d=4,5$). We have incorporated this in the phase diagram in Figure \ref{phase1}
for the case $d=5$ by using the data points found by Wiseman in
\cite{Wiseman:2002zc}. We refer to Appendix \ref{appwis}
for details on this numerically obtained solution.

We thus see that all the known solutions corresponding to black
holes or black strings can be mapped in our $(M,n)$ phase diagram,
as done in the phase diagram of Figure \ref{phase1}.

\section{Thermodynamics in the $(M,n)$ phase diagram}
\label{secbranch}

In this section we first derive a new Smarr formula
for static and neutral black holes/strings on cylinders.
We then show how one can use this to study the thermodynamics
of branches of solutions in the $(M,n)$ phase diagram.

To derive the new Smarr formula, we begin by considering the
Komar integral
\begin{equation}
I_S = - \frac{1}{16 \pi G_N} \int_S dS_{\mu \nu} D^\mu k^\nu \ ,
\end{equation}
where $S$ is a $d-1$ dimensional hypersurface in the $d+1$ dimensional
cylinder space-time and $k$ is a killing vector for the metric.
We can also write this integral as
\begin{equation}
I_S = - \frac{1}{16 \pi G_N} \int_S \sqrt{-g} \frac{1}{(d-1)!}
\epsilon_{\mu \nu \kappa^1 \cdots \kappa^{d-1}} D^\mu k^\nu
\, dx^{\kappa_1} \wedge \cdots \wedge dx^{\kappa_{d-1}} \ .
\end{equation}

Consider now a static neutral solution on the cylinder with an event horizon.
Consider furthermore a certain time $t=t_0$.
Define $S_h$ to be the null-surface of the event horizon at $t=t_0$.
We also choose a
$d-1$ dimensional surface at $r=\infty$ for $t=t_0$,
which we call $S_\infty$, so that essentially $S_\infty = S^{d-2} \times S^1$.
By Gauss theorem we have
\begin{equation}
I_{S_h} -I_{S_{\infty}}
= \frac{1}{16 \pi G_N} \left( \int_{S_{\infty}} dS_{\mu \nu} D^\mu k^\nu
- \int_{S_h} dS_{\mu \nu} D^\mu k^\nu \right)
= \frac{1}{8 \pi G_N} \int_{V} dS_\mu D_\nu D^\mu k^\nu \ ,
\end{equation}
where $V$ is the $d$ dimensional volume between $S_h$ and $S_\infty$
so that $\partial V = S_h \cup S_\infty$.
Since for a killing vector we have $D_\nu D^\mu k^\nu
= R^\mu_{\ \nu} k^\nu$ and since $R_{\mu \nu} = 0$ everywhere in $V$,
as we are away from the black hole/string, it follows that $D_\nu D^\mu
k^\nu = 0$. Thus, we find that $I_{S_h} = I_{S_{\infty}}$.

Since we have a static solution we can choose $k$ to be the
time-translation killing vector field, i.e. $k = \partial / \partial t$.
On the event horizon $S_h$ we can choose a normal $n$ such that
$k_\mu n^\mu = -1$. Then one can show that
$dS_{\mu \nu} = (k_\mu n_\nu - k_\nu n_\mu) dA $ where
$dA$ is the area element on the event horizon (see \cite{Townsend:1997ku}).
Using this, we have \cite{Townsend:1997ku}
\begin{equation}
\label{ISH}
I_{S_h} = - \frac{1}{8 \pi G_N} \int_{S_h} dA k^\nu ( D_\nu k^\mu ) n_\mu
= - \frac{\kappa }{8 \pi G_N} \int_{S_h} dA k^\mu n_\mu
= \frac{\kappa A}{8 \pi G_N} = T S \ ,
\end{equation}
where we used that $k^\nu D_\nu k^\mu = \kappa \, k^\mu$.
Here $\kappa $ is the surface gravity, $k_\mu n^\mu = -1$ and
we used the thermodynamic identifications
$\kappa = 2\pi T$ and $A = 4 G_N S$.
On the other hand, we have asymptotically
\begin{equation}
\label{ISI}
I_{S_{\infty}} =
- \frac{1}{16 \pi G_N} \int_{S_{\infty}} dS_{0r} \partial_r g_{00}
= \frac{\Omega_{d-2} 2\pi R_T}{16 \pi G_N} (d-3) c_t
= \frac{d-2-n}{d-1} M \ ,
\end{equation}
where we first used that
 the non-zero components of $D^\mu k^\nu$ are
$D^0 k^r = - D^r k^0 = \frac{1}{2} \partial_r g_{00}$ to leading order,
and \eqref{findMn} in the last step.

In conclusion, we get from \eqref{ISH} and \eqref{ISI} that
\begin{equation}
\label{smarrform}
TS = \frac{d-2-n}{d-1} M \ .
\end{equation}
This is a new Smarr formula which holds for any static and neutral
black hole/string on a cylinder.
Indeed the Smarr formula \eqref{smarrform}
can be checked for all three known branches in the phase diagram.
For uniform black strings ($n=1/(d-2)$) as well as for small black holes on
the cylinder ($n=0$) it reduces to known Smarr relations.
In Appendix \ref{appwis} we show that the
formula also holds to very high precision
for the non-uniform black string branch using the numerical data of Wiseman.

Define now
a {\sl branch} to be a series of solutions corresponding
to a curve in the $(M,n)$ diagram for which the first law
of thermodynamics $\delta M = T \delta S$ is obeyed when moving between the
solutions.%
\footnote{We show in \cite{Harmark:2003eg} that {\sl all}
series of solutions which are connected by infinitesimal steps
(i.e. being differentiable)  have the property  that the first law
$\delta M = T \delta S$ is always obeyed. In other words, for a
smooth curve in the $(M,n)$ phase diagram the first law
is obeyed on the curve.}
Then if we consider a branch for which we
know the precise curve in the $(M,n)$ phase diagram, we can find
the thermodynamics on the branch. For simplicity we consider here
the case where the curve is given by a function $n =
n(M)$. Using now $\delta M = T \delta S$ and the Smarr formula
\eqref{smarrform} we
get
\begin{equation}
\label{intS}
\frac{\delta \log S}{\delta M} = \frac{1}{M} \frac{d-1}{d-2-n(M)} \ .
\end{equation}
Thus, we see we only need to know the entropy in one point of the
branch to find the complete function $S(M)$ for the branch.%
\footnote{Notice however
the important subtlety that if the point in which one knows the entropy has
zero entropy then more information is needed to get the entropy function
$S(M)$ since it is evident from \eqref{intS} that one can only
get $\log S (M)$ up to an additive constant in that case.}
A similar statement applies to the temperature.
In conclusion, knowing the exact curve of a branch in the $(M,n)$
phase diagram enables one to obtain the entire thermodynamics of that branch.

Another useful feature of depicting a branch in the
$(M,n)$ phase diagram is that
when two branches intersect we have the following Intersection Rule
stating which branch has the highest entropy.%
\footnote{Notice however that if the branches intersect in a point for
which the entropy is zero the Intersection Rule is not necessarily true.
If we denote $M_\star$ the mass corresponding to the intersection
then we need that $S_1(M)/S_2(M) \rightarrow 1$ for $M \rightarrow M_\star$
in order for the Intersection Rule to hold.}
The proof of the Intersection Rule is at the end of this section.

\begin{itemize}
\item[$\blacktriangleleft$]  {\bf Intersection Rule} \newline
Consider two branches
$\gamma_{1,2}$ in the $(M,n)$ phase diagram,
parameterized by the functions $n_1(M)$ and $n_2(M)$.
Let the range of masses be from $M'$ to $M''$ ($M' < M''$)
and assume that $n_1(M) > n_2(M)$ for $M' < M < M''$.
Then
\begin{equation}
\label{lem1}
\gamma_1 \mbox{ intersect } \gamma_2 \mbox{ at } M=M' \ \Rightarrow \
S_1 (M) > S_2 (M) \mbox{  for  } M > M'  \ ,
\end{equation}
and
\begin{equation}
\label{lem2}
\gamma_1 \mbox{ intersect } \gamma_2 \mbox{ at } M=M'' \ \Rightarrow \
S_1 (M) < S_2 (M) \mbox{  for  } M < M''  \ .
\end{equation}
\end{itemize}

\begin{figure}[ht]
\centerline{\epsfig{file=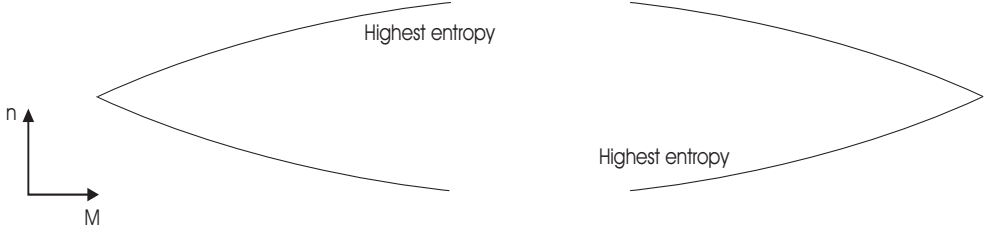,width=13cm,height=3cm}}
\caption{Illustration of the Intersection Rule.}
\label{figpeda}
\end{figure}

Thus, we see that for two
intersecting branches we have the property
that for masses below an intersection point the branch
with the lower relative binding energy has the highest entropy,
whereas for masses above an intersection point it is the branch
with the higher relative binding energy that has the highest entropy.
We have illustrated this in Figure \ref{figpeda}.

As a consequence of the Intersection Rule,
we find that it is impossible
for two branches which are described by well-defined functions
$n=n(M)$ to intersect twice.
If we write the two functions corresponding
to the two branches as $n_1(M)$ and $n_2(M)$ and the
two intersections as $M'$ and $M''$, then the
Intersection Rule gives a clear contradiction since we get from
\eqref{lem1} that $S_1 (M'') > S_2 (M'')$ and from
\eqref{lem2} that $S_1 (M') < S_2 (M')$ both of which contradict
the fact that the two branches intersect in $M'$ and $M''$.

We remind the reader of the physical reason why
we are comparing the entropy of two branches
at a given value of $M$, i.e. in the
micro canonical ensemble. This is because,
in order to address the stability of a
particular branch, one would like to examine whether it is entropically
favorable for the solution to move to another branch, characterized
by a different value of the relative binding energy,
by redistributing its mass.

The proof of above stated Intersection Rule
goes as follows. Let $S_i(M) \equiv S(M,n_i(M))$,
$i = 1,2$ be the entropy on the two branches $\gamma_{1,2}$.
If the two branches meet at $M=M_\star$ we can write the relation
\begin{equation}
\label{S12}
\frac{S_1 (M)}{S_2(M)} = 1 + \int_{M_\star}^M \delta \tilde{M}
\frac{\delta}{\delta\tilde{M}} \left( \frac{S_1 (\tilde{M})}{S_2(\tilde{M})}
\right) \ .
\end{equation}
Obviously $M_\star$ is either $M'$ or $M''$.
Consider now the differential
\begin{equation}
\label{ddM}
\frac{\delta}{\delta M} \left( \frac{S_1 (M)}{S_2(M)} \right)
= \frac{T_2 S_2 - T_1 S_1}{T_1 T_2 S_2^2}
= (n_1 - n_2) \frac{M}{(d-1)T_1 T_2 S_2^2} \ ,
\end{equation}
where we used the first law of thermodynamics $\delta M_i = T_i \delta S_i$ for
each of the branches in the first step
and the Smarr formula \eqref{smarrform} in the second step.
Since we assume that $n_1(M) > n_2(M)$ for $M' < M < M''$
we have
\begin{equation}
\label{ddM2}
\frac{\delta}{\delta M} \left( \frac{S_1 (M)}{S_2(M)} \right) > 0
\mbox{ for } M' < M < M'' \ .
\end{equation}
Using \eqref{ddM2} along with \eqref{S12},
we see that putting $M_\star = M'$ gives \eqref{lem1} and
putting $M_\star = M''$ gives \eqref{lem2}, completing the proof.

\section{Virtues of the $(M,n)$ diagram and comparison
to the $(\lambda,M)$ diagram}
\label{secvirt}

So far in the literature, the phase diagram used to display the
various phases of strings and black holes on cylinders has been
what we call the $(\lambda,M)$ diagram. It is defined
using the so-called non-uniformity parameter of \cite{Gubser:2001ac}
\begin{equation}
\label{lamb}
\lambda = \frac{1}{2} \left( \frac{R_{\rm max}}{R_{\rm min}} - 1 \right) \ ,
\end{equation}
where $R_{\rm max}$ and $R_{\rm min}$ are defined respectively as
the maximal and the minimal radius of the $S^{d-2}$ sphere on
the horizon.
Note that $\lambda$ is invariant under coordinate transformations.%
\footnote{This can easily be seen by considering the most general
static metric with spherical symmetry in the $\R^{d-1}$ part.
Write here this metric as
$$
ds^2 = - U dt^2 + g_{ij} dx^i dx^j + V d\Omega_{d-2}^2 \ ,
$$
with $i,j=1,2$ and
with $U$, $V$ and $g_{ij}$ being functions only of $x^1$ and $x^2$.
Then the size of the $S^{d-2}$ sphere at a point on the horizon is
given by the value of $V$ at that point. It is then clear
that the maximum and minimum of $V$ is
invariant under any coordinate transformation
$(x^1,x^2) \rightarrow (\tilde{x}^1,\tilde{x}^2)$ since $V$ transform
as a scalar.}
In the following we discuss some
important differences between the $(\lambda,M)$
and the $(M,n)$ phase diagrams.

An immediate physical point to notice is that $M$ and $n$
completely characterize the leading correction to the space-time
at infinity.
Thus, if an observer at infinity should draw diagrams
of his/her findings, it would be natural to draw these in an $(M,n)$
phase diagram.
In contrast, $\lambda$ can not be measured at infinity, since to know
$\lambda$ one has to know the space-time all the way to the horizon.

Note also that the parameter $n$ always stays
finite contrary to $\lambda$ which is infinite
for the black hole branch since $R_{\rm min} = 0$ on this branch.
Moreover, we have already noted in Section \ref{secdefMn} that
$0 \leq n \leq d-2$.
However, we do not expect $n > 1/(d-2)$ since
it seems reasonable to assume that the uniform string branch,
which has $n=1/(d-2)$, has the maximal possible relative binding
energy.
We therefore have the bound $0 \leq n \leq \frac{1}{d-2} $.
That $n$ is bounded makes it useful as a measure for
``how far away'' a given string solution is from a given black hole
solution.

However, we should note that the $(\lambda,M)$ diagram has
information that the $(M,n)$ diagram does not: The value of
$\lambda$ explicitly states whether a solution is a uniform string,
a non-uniform string or a black hole.
On the other hand, as we described in Section \ref{secbranch},
the $(M,n)$ phase diagram has several useful features.
For example we have explained how one can
read off the thermodynamics of a branch of solutions using
the $(M,n)$ diagram. Moreover, it was explained that there are certain
rules in the $(M,n)$ diagram restricting which branches are possible.

\section{Discussion and conclusions}
\label{secconcl}

We have discussed in this paper a novel type of phase diagram for
black holes and black strings on cylinders. The phase diagram
involves, beyond the mass $M$, a new asymptotic quantity $n$,
which is the relative binding energy in the circle direction.
So far, three branches of solutions are known in the phase diagram:
the uniform black string, the non-uniform string solutions numerically obtained by Wiseman \cite{Wiseman:2002zc}
(see also \cite{Gubser:2001ac}) and the small black hole branch.
We have plotted these known branches in the new phase diagram.

We have  derived a new
Smarr formula, valid for black holes and strings on the cylinder,
which involves the quantity $n$. This Smarr formula provides
a powerful instrument as it enables one to obtain the entire
thermodynamics, once the solution $n(M)$ in the phase diagram is known.
Moreover, when combined with the first law of thermodynamics,
the Smarr formula also implies an Intersection Rule, which easily
determines the branch of highest entropy for intersecting  branches in the phase diagram.
Although we have focussed in this paper
on neutral black holes/strings, this part of the development is easily
generalized to the non-extremal and near-extremal case as well as
the inclusion of angular momenta.
For example, in the non-extremal case, the Smarr formula is simply
obtained by $M \rightarrow M - \mu Q$ in \eqref{smarrform}.

We also remark that the formalism introduced in this paper seems
well-suited for the important study of the phase structure of
black $p$-branes in spaces with more than one compact direction.
The simplest generalization one can think of would be to study
black holes, strings and membranes on the torus $\R^{d-2} \times T^2$.
Following our development, one would need in that case, beyond
the mass $M$, three additional asymptotic quantities corresponding
to the stress in the torus direction.

We have pointed out some of the virtues of the $(M,n)$ phase diagram
and compared to the $(\lambda,M)$ diagram, so far used in the literature.
In a forthcoming work \cite{Harmark:2003eg} 
we will use the formalism introduced
in this paper to obtain various new insights into the phase
structure for black holes and strings on cylinders.

Finally, we comment on the issue of uniqueness of black hole/string
solutions on a cylinder. We obtained in Section \ref{secdefMn}
the gauge invariant quantities that determine the leading
order metric in the presence of a static and neutral (spherical symmetric)
mass distribution on the cylinder.
Beyond the mass $M$ and the
relative binding energy $n$, these were shown to also involve the higher
Fourier modes $\varrho_{k \geq 1}$ of the mass distribution.
For black holes/strings on  cylinders this implies the possibility
of having several solutions for a given point in the $(M,n)$ phase
diagram, distinguished by the $\varrho_k$.

For neutral black holes in
four space time dimensions it is well-known that the solution is
uniquely characterized given the mass.%
\footnote{For asymptotically flat  solutions in higher dimensions
no simple uniqueness theorems are presently known. In particular,
in Ref.~\cite{Emparan:2001wn} a new black ring solution was found in
five dimensions.}
It would naturally be very interesting
to examine whether similar uniqueness theorems
hold for black holes/strings on cylinders.%
\footnote{We recall here that our analysis pertains to
cylinders $\R^{d-1} \times S^1$, $d \geq 4$, so that the
number of space time dimensions is five or higher.}
For example, it may  be that
the Fourier modes $\varrho_k$ are determined in terms of $M$ and $n$.
It is thus conceivable that the following {\it uniqueness hypothesis}
is valid:
\begin{itemize}
\item[$\blacksquare$] Consider solutions of the Einstein equations on a
cylinder $\R^{d-1} \times S^1$.
For a given mass $M$ and relative binding energy $n$
there exists
at most one neutral and static solution of the Einstein equations
with an event horizon.
\end{itemize}
A less strong version of the above hypothesis would be that also
the choice of horizon topology ($S^{d-2} \times S^1$ or $S^{d-1}$)
needs to be specified in order to obtain a unique solution.%
\footnote{Note that
in \cite{Kol:2002dr} a different proposal for a uniqueness theorem
was given: For a given mass $M$ there exists only one
stable solution of a certain topology.
That proposal is of a rather different nature, since if correct
it does not address the point of having different branches that
exists classically.
 Whether this proposal is correct or not is
thus independent of the above discussion of the uniqueness
properties relating to $M$ and $n$.}

\section*{Acknowledgments}

We thank J. de Boer, G. Horowitz, V. Hubeny, M. Rangamani, S.
Ross, K. Skenderis, M. Taylor and E. Verlinde for helpful
discussions. We especially thank Toby Wiseman for many valuable
discussions and comments, and for providing us with many details
and explanations on the numerical solutions of
\cite{Wiseman:2002zc}.

\begin{appendix}

\section{Green function on a cylinder}
\label{appgreen}

We consider the cylinder $\R^{d-1} \times S^1$ with metric
\begin{equation}
ds^2 = dz^2 + dr^2 + r^2 d\Omega_{d-2}^2 \ .
\end{equation}
where $z$ is periodic with period $2\pi R_T$.
Here the $S^{d-2}$ sphere is parameterized by
the angles $\theta_1,...,\theta_{d-2}$ and it has the metric
\begin{equation}
d\Omega_{d-2}^2 = \sum_{a,b=1}^{d-2} \gamma_{ab} \, d\theta_a d\theta_b \ .
\end{equation}
Define $\gamma = \sqrt{ \det ( \gamma_{ab} ) }$.
Define furthermore the notation $\Theta = (\theta_1,...\theta_{d-2})$.

We define then the Green function on $\R^{d-1} \times S^1$ as
\begin{equation}
\label{greendef}
\nabla_{z,r,\Theta}^2 G (z-z',r-r',\Theta-\Theta') = -
\frac{\Omega_{d-2}}{r^{d-2} \gamma}
\delta (z-z') \delta (r-r') \delta (\Theta - \Theta') \ .
\end{equation}
In the following we derive a useful expansion of the Green function
on the cylinder which is used in Section \ref{secdefMn}.

We first introduce the spherical harmonics $Y_l (\Theta)$ which
are eigenfunctions
\begin{equation}
\frac{1}{\gamma} \sum_{a=1}^{d-2} \partial_{\theta_a}
\left[ \gamma \gamma^{ab} \partial_{\theta_b} Y_l (\theta_1,...,\theta_{d-2}) \right]
= - h_l Y_l(\theta_1,...,\theta_{d-2}) \ ,
\end{equation}
that satisfy the completeness relation
\begin{equation}
\sum_l
Y^{*}_l (\Theta') Y_l (\Theta)
= \frac{1}{\gamma} \delta (\Theta - \Theta') \ .
\end{equation}
We can now expand any function on $S^{d-2}$ as
$f(\Theta) = \sum_l a_l Y_l(\Theta)$.
We can furthermore expand the delta function for $z$ as
\begin{equation}
\delta ( z - z' ) = \frac{1}{2\pi R_T} \sum_{k \in \Z} e^{ik(z-z')/R_T}
=  \frac{1}{2\pi R_T} \left[ 1 + 2 \sum_{k=1}^\infty
\cos \! \left( \frac{k(z-z')}{R_T} \right) \right] \ .
\end{equation}
The Green function can then be expanded as
\begin{equation}
G(z-z',r-r',\Theta-\Theta') = \frac{1}{2\pi R_T}
\sum_l Y^*_l (\Theta') Y_l (\Theta)
\sum_{k \in \Z} M^{(k)}_l ( r,r') e^{ik(z-z')/R_T} \ .
\end{equation}
Using this the Green function equation \eqref{greendef}
then becomes equivalent to
\begin{equation}
\label{Mnl}
\left( \partial_r^2 + \frac{d-2}{r} \partial_r \right)
M^{(k)}_l (r,r')
- \left( \frac{k^2}{R_T^2} + \frac{h_l}{r^2}  \right) M^{(k)}_l (r,r')
= - \frac{\Omega_{d-2}}{r^{d-2}} \delta (r-r') \ .
\end{equation}
In order to solve this equation, we notice that
the modified Bessel functions of the second kind $K_\nu(x)$ and $I_\nu(x)$
are solutions to the equation
\begin{equation}
f'' + \frac{1}{x} f' - \left( 1 + \frac{\nu^2}{x^2} \right) f = 0 \ .
\end{equation}
Notice then that if we define $g$ as
\begin{equation}
g(x) = x^{-\frac{d-3}{2}} f( kx ) \ ,
\end{equation}
then $g$ is a solution to the equation
\begin{equation}
\label{gde}
g'' + \frac{d-2}{x} g'
- \left[ k^2 + \left( \nu^2 - \frac{1}{4}(d-3)^2 \right) \right] g = 0 \ .
\end{equation}
So, using
\begin{equation}
\nu_l = \sqrt{ \frac{1}{4}(d-3)^2  + h_l } \ ,
\end{equation}
we see that $g(r/R_T)$ is a solution of \eqref{Mnl} when $r \neq r'$.
Define now the two independent solutions of the differential equation
\eqref{gde} by
\begin{equation}
\psi_1^{(k)} (\nu,x) = k^{-\nu} x^{-\frac{d-3}{2}} I_\nu (kx)
\spa
\psi_2^{(k)} (\nu,x) = k^{\nu} x^{-\frac{d-3}{2}}  K_\nu (kx) \ ,
\end{equation}
satisfying
\begin{equation}
\psi_1^{(k)} (\nu,x) \simeq 
\frac{1}{2^\nu \Gamma (\nu +1)} x^{\nu - \frac{d-3}{2}} \spa
\psi_2^{(k)} (\nu,x) \simeq 
2^{\nu-1} \Gamma (\nu ) x^{-\nu - \frac{d-3}{2}} \spa
x \ll 1 
\end{equation}
\begin{equation}
\psi_1^{(k)} \frac{d}{dx} \psi_2^{(k)} -
\psi_2^{(k)} \frac{d}{dx} \psi_1^{(k)} = -\frac{1}{x^{d-2}}  
 \ .
\end{equation} 
Then we can write the solution of \eqref{Mnl} as
\begin{equation}
M^{(k)}_l (r,r') = \frac{\Omega_{d-2}}{R_T^{d-3}} \,
\psi_1^{(k)} \! \left( \nu_l, \frac{r_<}{R_T} \right)
\psi_2^{(k)} \! \left(\nu_l, \frac{r_>}{R_T} \right) \ ,
\end{equation}
with
\begin{equation}
r_< = \left\{ \begin{array}{ll} r \ , & \ r < r' \\ r' \ , & \ r > r'
\end{array} \right.
 \spa r_> = \left\{ \begin{array}{ll} r' \ , & \ r < r' \\ r \ , & \ r > r'
\end{array} \right. \ \ .
\end{equation}

\section{Critical masses for the Gregory-Laflamme instability}
\label{appMGL}

For completeness we list in this appendix
the Gregory-Laflamme masses for uniform black strings
on a cylinder $\R^{d-1} \times S^1$ with $d=4,...,9$.
The data has been taken from \cite{Gregory:1993vy} (the published
version) by reading off $\mu$ from their Fig. 1.
The values of
$\mu$ that we read off are $0.44, 0.62, 0.80,0.93,1.04,1.15$
for $d=4,...,9$. One can then see in
\cite{Gregory:1993vy,Gregory:1994bj}
that the critical horizon radius is given by $r_0/R_T = 2\mu$, where
$r_0$ is the horizon radius of the uniform black string solution
\eqref{metunstr} and $R_T$ is the radius of the cylinder.
Using that
\begin{equation}
M = \frac{\Omega_{d-2} 2\pi R_T}{16 \pi G_N} (d-2) r_0^{d-3} \ ,
\end{equation}
we can compute the critical masses. In Table \ref{tabMGL} we have
listed the horizon radii and the masses for $d=4,...,9$.
We have used the notation that the circumference of the cylinder
is $L = 2\pi R_T$.

\begin{table}[h]
\begin{center}
\begin{tabular}{|c||c|c|c|c|c|c|}
\hline
$d$ & $4$ & $5$ & $6$ & $7$ & $8$ & $9$
\\ \hline
$r_0/L$ & $0.14$ &$0.20$ &$0.25$ &$0.30$ &$0.33$ &$0.37$
\\ \hline
$16\pi G_N M_{\rm GL}/L^{d-2}$ & $3.52$ & $2.31$ & $1.74$ & $1.19$ & $0.79$ & $0.55$ \\
\hline
\end{tabular}
\caption{Critical horizon radii and masses for the
Gregory-Laflamme instability.
 \label{tabMGL}}
\end{center}
\end{table}

\section{Details on Wiseman's non-uniform black string data}
\label{appwis}

In this appendix we provide the reader with details of how the
plot of Wiseman's non-uniform black string branch in Figure
\ref{phase1} was made. Wiseman's non-uniform black string
solutions (with $d=5$) were obtained in \cite{Wiseman:2002zc} by
numerically solving Einstein's equations for the
conformal ansatz
\begin{equation}
\label{Wise}
ds_6^2 = -\frac{r^2}{m+ r^2} e^{2 A} dt^2 + e^{2B} (dr^2 + dz^2)
+ e^{2C} (m +r^2) d \Omega_3^2 \ .
\end{equation}
Here, $A,B,C$ are functions of $(r,z)$, with asymptotic forms
\begin{equation}
\label{Wisebc}
A \simeq \frac{a_2}{r^2} \spa B \simeq \frac{b_2}{r^2} \spa C \simeq \frac{c_1}{r}
+ \frac{c_2}{r^2} \ .
\end{equation}

\begin{figure}[ht]
\centerline{\epsfig{file=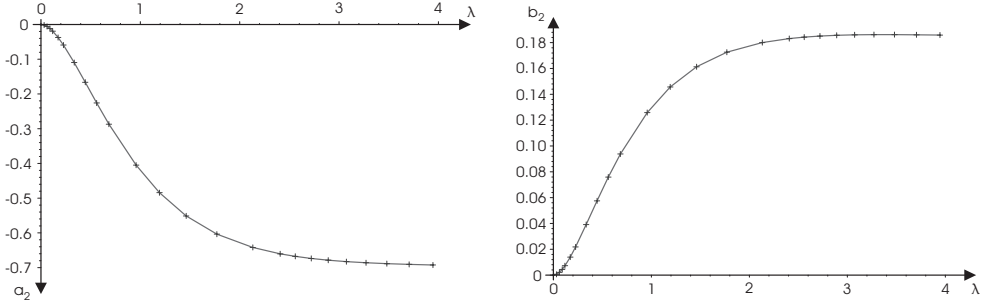,width=13cm,height=4cm}}
\caption{\small Plot of $a_2$ and $b_2$ versus $\lambda$ using
the data of \cite{Wiseman:2002zc}.}
\label{plotab}
\end{figure}

The data found in \cite{Wiseman:2002zc}
consist of 25 points, each corresponding to a non-uniform black string
solution. Each of the 25 points is then specified
by six numbers: $\lambda$, $a_2$, $b_2$, $T_{\rm w}$, $S_{\rm w}$,
and $M_{\rm w}$. $\lambda$ is defined in \eqref{lamb},
$T_{\rm w}$, $S_{\rm w}$ and $M_{\rm w}$ are the temperature,
entropy and mass that was computed in \cite{Wiseman:2002zc},
and $a_2$ and $b_2$ are defined in \eqref{Wisebc}.
These data was generously made available to us by Toby Wiseman.

\begin{figure}[ht]
\centerline{\epsfig{file=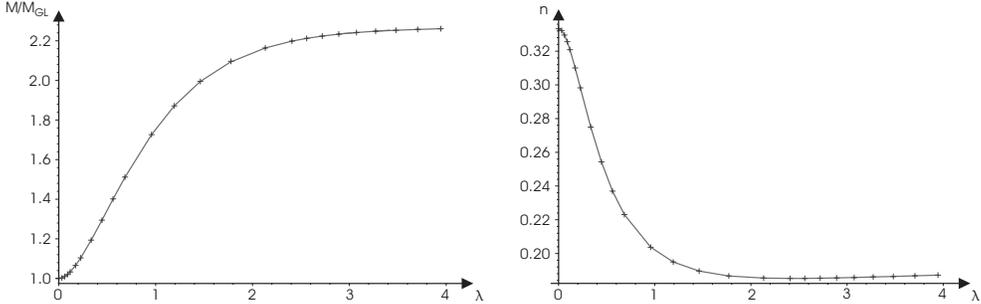,width=13cm,height=4cm}}
\caption{\small Plot of $M$ and $n$ versus $\lambda$.
$M$ and $n$ are computed from the $a_2$ and $b_2$ data.
The crosses mark the actual data-points.}
\label{plotMnlamb}
\end{figure}

\begin{figure}[ht]
\centerline{\epsfig{file=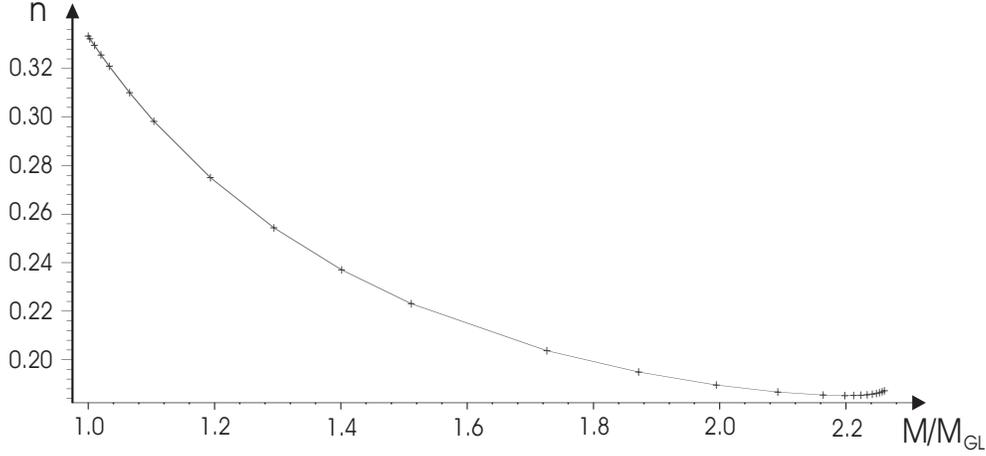,width=13cm,height=6cm}}
\caption{\small Plot of $M$ versus $n$ as computed
from the $a_2$ and $b_2$ data.
The crosses mark the actual data-points.}
\label{plotMn}
\end{figure}

The only data necessary to make the phase diagram in Figure \ref{phase1}
are $a_2$ and $b_2$ for the 25 points.
We have plotted $a_2$ and $b_2$ versus $\lambda$ in Figure \ref{plotab}.
We refer the reader to \cite{Wiseman:2002zc} for plots
of $T_{\rm w}$, $S_{\rm w}$ and $M_{\rm w}$ versus lambda.
By comparing \eqref{ctcz} and \eqref{Wise}-\eqref{Wisebc}
we see that $a_2$ and $b_2$
are related to $c_t$ and $c_z$ as
\begin{equation}
c_t = 1  - 2 a_2 \spa c_z = 2 b_2 \ ,
\end{equation}
where we note that $m=1$ in \eqref{Wise}.
We can then use this in \eqref{findMn} (with $d=5$) to determine the
mass $M$ and relative binding energy $n$ for each data point according to
\begin{equation}
\frac{M}{M_{\rm GL}}
= c_t - \frac{c_z}{3} \spa n = \frac{c_t -3c_z}{3 c_t -c_z} \ .
\end{equation}
In Figure \ref{plotMnlamb} we have plotted $M$ and $n$ versus $\lambda$.
Moreover, in Figure \ref{plotMn} we have plotted $M$ versus $n$.

\begin{figure}[ht]
\centerline{\epsfig{file=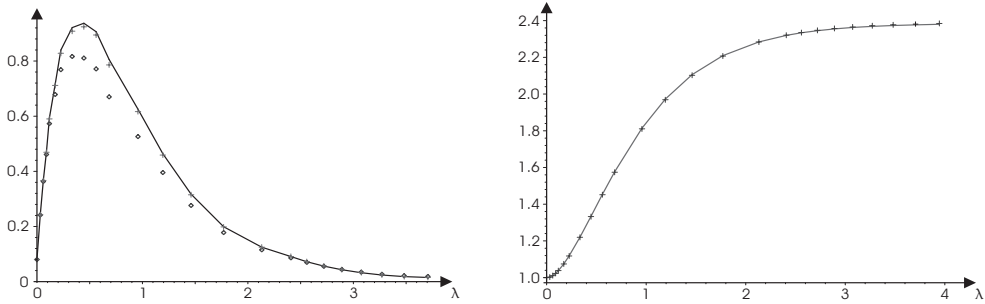,width=13cm,height=4cm}}
\caption{\small On the left plot we have checked the
first law of thermodynamics for Wiseman's branch of solutions.
Here the solid curve shows $T \frac{\delta S}{\delta \lambda}$, the
crosses show $\frac{\delta M}{\delta \lambda}$
and the diamonds show $\frac{\delta M_w}{\delta \lambda}$, all
in units of $M_{GL}$. On the right plot we have checked Smarr's formula
\eqref{smarrform}. Here the solid curve show $TS$ while
the crosses show $\frac{3-n}{4}M$, both in units of
$T_{\rm GL} S_{\rm GL} = \frac{2}{3}M_{\rm GL}$.}
\label{plotfirstlaw}
\end{figure}

We now check that the first law of thermodynamics $\delta M = T \delta S$
is obeyed for the above computed $M$.
In left graph of
Figure \ref{plotfirstlaw} we have depicted $\delta M/\delta \lambda$
and $T \delta S / \delta \lambda$ versus $\lambda$.
We note that they agree to a high precision.
The third curve in Figure \ref{plotfirstlaw} is $\delta M_w/\delta \lambda$
where $M_w$ is the mass that Wiseman computed in the original version of
\cite{Wiseman:2002zc}. We see that the systematic error that Wiseman
noted in the original version of \cite{Wiseman:2002zc} in fact is due
to an erroneous computation of the mass, and not an inaccuracy of
the solutions that were found (see also comment in new version of
\cite{Wiseman:2002zc}).

In the right graph of Figure \ref{plotfirstlaw}
we have instead plotted $TS$ (computed from
$T_w$ and $S_w$) and $\frac{3-n}{4}M$ versus $\lambda$.
This is a check of Smarr's formula \eqref{smarrform}.
We see that Smarr's formula is correct to a very high precision.


\end{appendix}

\addcontentsline{toc}{section}{References}


\begin{thebibliography}{10}

\bibitem{Gregory:1993vy}
R.~Gregory and R.~Laflamme, ``Black strings and p-branes are unstable,'' {\em
  Phys. Rev. Lett.} {\bf 70} (1993) 2837--2840,
\href{http://arXiv.org/abs/hep-th/9301052}{{\tt hep-th/9301052}}.

\bibitem{Gregory:1994bj}
R.~Gregory and R.~Laflamme, ``The instability of charged black strings and
  p-branes,'' {\em Nucl. Phys.} {\bf B428} (1994) 399--434,
\href{http://arXiv.org/abs/hep-th/9404071}{{\tt hep-th/9404071}}.

\bibitem{Horowitz:2001cz}
G.~T. Horowitz and K.~Maeda, ``Fate of the black string instability,'' {\em
  Phys. Rev. Lett.} {\bf 87} (2001) 131301,
\href{http://arXiv.org/abs/hep-th/0105111}{{\tt hep-th/0105111}}.

\bibitem{Gubser:2001ac}
S.~S. Gubser, ``On non-uniform black branes,'' {\em Class. Quant. Grav.} {\bf
  19} (2002) 4825--4844,
\href{http://www.arXiv.org/abs/hep-th/0110193}{{\tt hep-th/0110193}}.

\bibitem{Wiseman:2002zc}
T.~Wiseman, ``Static axisymmetric vacuum solutions and non-uniform black
  strings,'' {\em Class. Quant. Grav.} {\bf 20} (2003) 1137--1176,
\href{http://www.arXiv.org/abs/hep-th/0209051}{{\tt hep-th/0209051}}.

\bibitem{Gregory:1988nb}
R.~Gregory and R.~Laflamme, ``Hypercylindrical black holes,'' {\em Phys. Rev.}
  {\bf D37} (1988)
305.

\bibitem{Choptuik:2003qd}
M.~W. Choptuik {\em et al.}, ``Towards the final fate of an unstable black
  string,'' {\em Phys. Rev.} {\bf D68} (2003) 044001,
\href{http://www.arXiv.org/abs/gr-qc/0304085}{{\tt gr-qc/0304085}}.

\bibitem{Harmark:2002tr}
T.~Harmark and N.~A. Obers, ``Black holes on cylinders,'' {\em JHEP} {\bf 05}
  (2002) 032,
\href{http://www.arXiv.org/abs/hep-th/0204047}{{\tt hep-th/0204047}}.

\bibitem{Horowitz:2002dc}
G.~T. Horowitz, ``Playing with black strings,''
\href{http://www.arXiv.org/abs/hep-th/0205069}{{\tt hep-th/0205069}}.

\bibitem{Kol:2002xz}
B.~Kol, ``Topology change in general relativity and the black-hole black-string
  transition,''
\href{http://www.arXiv.org/abs/hep-th/0206220}{{\tt hep-th/0206220}}.

\bibitem{Wiseman:2002ti}
T.~Wiseman, ``From black strings to black holes,'' {\em Class. Quant. Grav.}
  {\bf 20} (2003) 1177--1186,
\href{http://www.arXiv.org/abs/hep-th/0211028}{{\tt hep-th/0211028}}.

\bibitem{Harmark:2003fz}
T.~Harmark and N.~A. Obers, ``Black holes and black strings on cylinders,''
  {\em Fortsch. Phys.} {\bf 51} (2003) 793--798,
\href{http://www.arXiv.org/abs/hep-th/0301020}{{\tt hep-th/0301020}}.

\bibitem{Kol:2003ja}
B.~Kol and T.~Wiseman, ``Evidence that highly non-uniform black strings have a
  conical waist,'' {\em Class. Quant. Grav.} {\bf 20} (2003) 3493--3504,
\href{http://www.arXiv.org/abs/hep-th/0304070}{{\tt hep-th/0304070}}.

\bibitem{Casadio:2000py}
R.~Casadio and B.~Harms, ``Black hole evaporation and large extra dimensions,''
  {\em Phys. Lett.} {\bf B487} (2000) 209--214,
\href{http://www.arXiv.org/abs/hep-th/0004004}{{\tt hep-th/0004004}}.

\bibitem{Casadio:2001dc}
R.~Casadio and B.~Harms, ``Black hole evaporation and compact extra
  dimensions,'' {\em Phys. Rev.} {\bf D64} (2001) 024016,
\href{http://www.arXiv.org/abs/hep-th/0101154}{{\tt hep-th/0101154}}.

\bibitem{Horowitz:2002ym}
G.~T. Horowitz and K.~Maeda, ``Inhomogeneous near-extremal black branes,'' {\em
  Phys. Rev.} {\bf D65} (2002) 104028,
\href{http://www.arXiv.org/abs/hep-th/0201241}{{\tt hep-th/0201241}}.

\bibitem{DeSmet:2002fv}
P.-J. De~Smet, ``Black holes on cylinders are not algebraically special,'' {\em
  Class. Quant. Grav.} {\bf 19} (2002) 4877--4896,
\href{http://www.arXiv.org/abs/hep-th/0206106}{{\tt hep-th/0206106}}.

\bibitem{Kol:2002dr}
B.~Kol, ``Speculative generalization of black hole uniqueness to higher
  dimensions,''
\href{http://www.arXiv.org/abs/hep-th/0208056}{{\tt hep-th/0208056}}.

\bibitem{Sorkin:2002nu}
E.~Sorkin and T.~Piran, ``Initial data for black holes and black strings in
  5d,'' {\em Phys. Rev. Lett.} {\bf 90} (2003) 171301,
\href{http://www.arXiv.org/abs/hep-th/0211210}{{\tt hep-th/0211210}}.

\bibitem{Frolov:2003kd}
A.~V. Frolov and V.~P. Frolov, ``Black holes in a compactified spacetime,''
  {\em Phys. Rev.} {\bf D67} (2003) 124025,
\href{http://www.arXiv.org/abs/hep-th/0302085}{{\tt hep-th/0302085}}.

\bibitem{Emparan:2003sy}
R.~Emparan and R.~C. Myers, ``Instability of ultra-spinning black holes,'' {\em
  JHEP} {\bf 09} (2003) 025,
\href{http://www.arXiv.org/abs/hep-th/0308056}{{\tt hep-th/0308056}}.

\bibitem{Kol:2003if}
B.~Kol, E.~Sorkin, and T.~Piran, ``Caged black holes: Black holes in
  compactified spacetimes {I} -- theory,''
\href{http://www.arXiv.org/abs/hep-th/0309190}{{\tt hep-th/0309190}}.

\bibitem{Traschen:2001pb}
J.~Traschen and D.~Fox, ``Tension perturbations of black brane spacetimes,''
\href{http://www.arXiv.org/abs/gr-qc/0103106}{{\tt gr-qc/0103106}}.

\bibitem{Townsend:2001rg}
P.~K. Townsend and M.~Zamaklar, ``The first law of black brane mechanics,''
  {\em Class. Quant. Grav.} {\bf 18} (2001) 5269--5286,
\href{http://www.arXiv.org/abs/hep-th/0107228}{{\tt hep-th/0107228}}.

\bibitem{Traschen:2003jm}
J.~Traschen, ``A positivity theorem for gravitational tension in brane
  spacetimes,''
\href{http://www.arXiv.org/abs/hep-th/0308173}{{\tt hep-th/0308173}}.

\bibitem{Shiromizu:2003gc}
T.~Shiromizu, D.~Ida, and S.~Tomizawa, ``Kinematical bound in asymptotically
  translationally invariant spacetimes,''
\href{http://www.arXiv.org/abs/gr-qc/0309061}{{\tt gr-qc/0309061}}.

\bibitem{Witten:1981mf}
E.~Witten, ``A simple proof of the positive energy theorem,'' {\em Commun.
  Math. Phys.} {\bf 80} (1981)
381.

\bibitem{Gibbons:1983jg}
G.~W. Gibbons, S.~W. Hawking, G.~T. Horowitz, and M.~J. Perry, ``Positive mass
  theorems for black holes,'' {\em Commun. Math. Phys.} {\bf 88} (1983)
295.

\bibitem{Townsend:1997ku}
P.~K. Townsend, ``Black holes,''
\href{http://www.arXiv.org/abs/gr-qc/9707012}{{\tt gr-qc/9707012}}.

\bibitem{Harmark:2003eg}
T.~Harmark and N.~A. Obers, ``Phase structure of black holes and strings on
  cylinders,''
\href{http://www.arXiv.org/abs/hep-th/0309230}{{\tt hep-th/0309230}}.

\bibitem{Emparan:2001wn}
R.~Emparan and H.~S. Reall, ``A rotating black ring in five dimensions,'' {\em
  Phys. Rev. Lett.} {\bf 88} (2002) 101101,
\href{http://arXiv.org/abs/hep-th/0110260}{{\tt hep-th/0110260}}.

\end{thebibliography}

\providecommand{\href}[2]{#2}\begingroup\raggedright\endgroup

\end{document}